%% file: model.tex
\documentclass[letterpaper,10pt,conference]{ieeeconf}   
\usepackage{amsmath}    
\usepackage{amsfonts}
\usepackage{graphicx}   
\usepackage{epsfig} 
\usepackage{color}
\usepackage[normalem]{ulem}
\usepackage{cancel}
\usepackage{amssymb}
\usepackage{gensymb}
\usepackage{fixltx2e}
\usepackage{color}
\usepackage{array}
\usepackage{booktabs}

\title{\LARGE \bf
Building Model Identification during Regular Operation \\-- Empirical Results and Challenges}

\author{Qie Hu, Frauke Oldewurtel, Maximilian Balandat, Evangelos Vrettos, Datong Zhou, and Claire J. Tomlin
\thanks{This work is supported in part by NSF under CPS:ActionWebs (CNS-0931843) and CPS:FORCES (CNS1239166). The research of F. Olderwurtel has received funding from the European Union Seventh Framework Programme FP7-PEOPLE-2011-IOF under grant agreement number 302255, Marie Curie project `Stochastic Model Predictive Control, Energy Efficient Building Control, Smart Grid'.}
\thanks{Q. Hu, F. Oldewurtel, M. Balandat, D. Zhou and C.J. Tomlin are with the Department of EECS, 
University of California, Berkeley. \{qiehu,oldewurtel,balandat,datong.zhou,tomlin\}@berkeley.edu. E. Vrettos is with Power Systems Laboratory, ETH Z\"{u}rich. 
vrettos@eeh.ee.ethz.ch.}
}

\begin{document}
\maketitle
\thispagestyle{empty}
\pagestyle{empty}

\begin{abstract}
The inter-temporal consumption flexibility of commercial buildings can be harnessed to improve the energy efficiency of buildings, or to provide ancillary service to the power grid. To do so, a predictive model of the building's thermal dynamics is required. In this paper, we identify a physics-based model of a multi-purpose commercial building including its heating, ventilation and air conditioning system during regular operation. We present our empirical results and show that large uncertainties in internal heat gains, due to occupancy and equipment, present several challenges in utilizing the building model for long-term prediction. In addition, we show that by learning these uncertain loads online and dynamically updating the building model, prediction accuracy is improved significantly.
\end{abstract}

\input{introduction.tex}

\input{formulation.tex}
\input{results.tex}

\section{Conclusions}
We describe an approach to construct a physics-based model of a multi-zone commercial building, which uses experimental data measured during regular building operation. 
We show that large uncertainties in internal gains present several challenges in applying the model for long term prediction of a building's thermal dynamics. 
In addition, we show that by dynamically updating the estimates of internal gains, the model's prediction accuracy is improved significantly. 

Future work will investigate the trade-off between uncertainty in internal gains and prediction accuracy, as well as the necessary model complexity for a good control performance, in particular for harnessing building flexibility. 
In addition, we are working on experimentally verifying the performance of controllers designed using our building model.

\section*{Acknowledgment}
The authors thank Rongxin Yin for providing the EnergyPlus model, David Sturzenegger for his assistance with using the BRCM Toolbox and the SDH building manager Domenico Caramagno for his help in facilitating our experiments. 



 \bibliographystyle{IEEEtran}
 \bibliography{references}
\end{document}

%% file: introduction.tex
\section{Introduction}

Commercial buildings account for more than 35\% of electricity consumption in the U.S., 39\% of which is due to heating, ventilation and air conditioning (HVAC) systems \cite{EnergyOutlook}. 
Energy consumption of HVAC systems can be partly shifted in time without compromising occupant comfort, because of buildings' inherent thermal capacity. 
As a result, there has been extensive research, using frameworks such as model predictive control (MPC), trying to harness this inter-temporal consumption flexibility and minimize energy usage of buildings \cite{Parisio:2014aa, Siroky:2011aa}. 
More recently, the feasibility of using buildings to provide ancillary services, such as frequency regulation, to the power grid has also been studied \cite{Baccino, Mehdi, Vrettos, Vrettos2016, Max}. 
Such applications require accurate models describing the thermal dynamic behavior of the buildings.

\subsection{Desired Model Features and Challenges}

The building models should be identified using actual experimental data and capture realistic disturbances, such as internal gains. 
Furthermore, bilinear, multi-zone models may quantify buildings' electricity consumption flexibility more precisely: the bilinear thermal dynamics naturally arise from the physics of the HVAC system (refer to Section \ref{sec:model} for details); controllers designed using multi-zone models can allow the temperature of a room to fluctuate when it is unoccupied, for instance, hence achieving energy savings which are not possible when simplified models that approximate the building as a single zone are used.

However, there are many challenges in identifying such a building model. 
First, actual buildings often have different types of spaces that are subject to very different uncertainties, e.g. occupancy, which are difficult to capture. 
Second, buildings are often not sufficiently excited, as they must satisfy strict regulatory requirements during regular operation, which limit the type and duration of excitation experiments that can be conducted \cite{Aswani}.  

To circumvent these difficulties, various approaches have been taken in the research community. 
In \cite{Aswani, Aswani_lbmpc, zhou2016}, the authors identify data-driven linear models for a single type of building space.
Lin \textit{et al.}  \cite{LinExp} conduct frequency regulation experiments under a controlled environment without disturbances such as solar radiation and occupants, thus simplifying the building model required to design the controller. 
Faced with insufficient excitation of buildings, Mehdi \textit{et al.} \cite{Mahdi} reduce the required model complexity by carrying out their experiments in a single room, whereas others \cite{Frauke, Ma_cooltower, Hao_lumped}
use lumped thermal models that approximate the building as a single zone, with an average building temperature. 
Finally, the authors of \cite{David, Sun} identify and test multi-zone models for a single floor and an entire building, respectively, using simulated data where uncertainties are precisely controlled or removed, and arbitrary excitations can be simulated.

These approaches are valuable in providing estimates of a building's consumption flexibility, however none of them delivers a model that satisfies all the aforementioned desired features for a more precise quantification of the building's flexibility.

\subsection{Contributions}

In this paper, we identify a semi-parametric model for a multi-zone commercial building during regular operation. Our main contributions are the following:
\begin{itemize}
\item
We propose a procedure to identify a physics-based model of a multi-zone building, that is easy to implement with the building in regular operation, and captures internal gains such as occupancy.
This procedure uses excitation experiments that actively perturb the building and generate data that can be used for more accurate parameter identification.
\item
We provide an analysis of the model's prediction accuracy versus its prediction horizon, and show that this model is limited in making long-term predictions, partly due to large disturbances such as internal gains, which are uncertain.
\item
Finally, we propose to dynamically update the estimate of internal gains 
based on current temperature measurements, and show that this significantly improves the building model's prediction accuracy.
\end{itemize}

\vspace*{0.2cm}
This paper is organized as follows: We begin by describing the building and the excitation experiments in Section \ref{sec:building}. We then present the building model in Section \ref{sec:model}. Section \ref{sec:model_id} describes our identification method for a fixed model and reports the prediction results. A model that is updated online is presented in Section \ref{sec:online_ig}. Finally, we provide a discussion of uses and challenges related to building models.

\vspace*{0.2cm}
\textit{Notation}: Unless stated otherwise, subscripts in italics as in $u_i$ denote instances of variables. Upright subscripts as in $U_\text{win}$ denote variable names.

%% file: formulation.tex
\section{Building and Excitation Experiments}\label{sec:building}

\subsection{Building}

\begin{figure}
	\center
	\includegraphics[width=1\columnwidth]{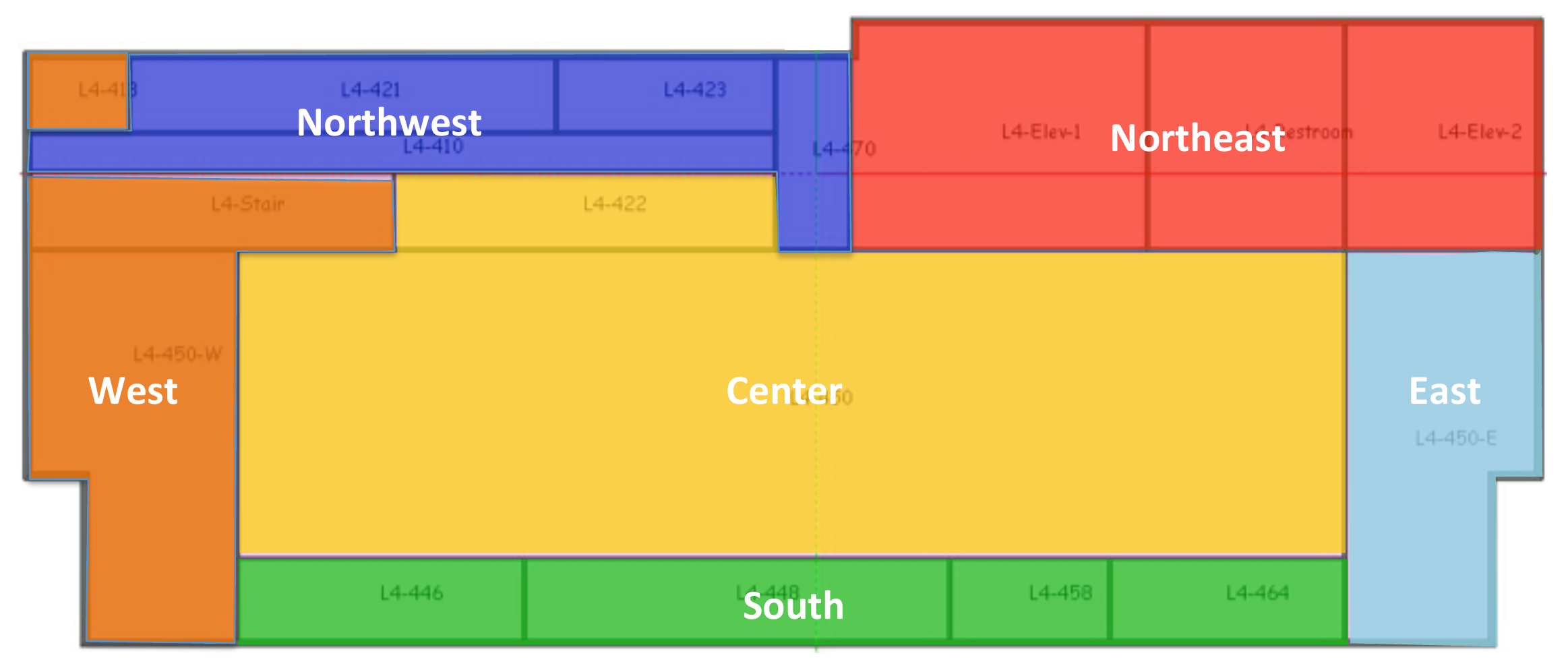}
	\caption{Zones for the 4th floor of Sutardja Dai Hall (SDH).}
	\label{Fig:floorplan}
\end{figure}

Sutardja Dai Hall (SDH) is a building located on the University of California, Berkeley campus. For ease of presentation, we focus on the entire 4th floor of SDH, which has a total floor area of 1300m$^2$. As shown in Figure \ref{Fig:floorplan}, we aggregate the rooms into 6 zones: Northwest (NW), West (W), South (S), East (E), Northeast (NE) and Center (C). The north-side rooms are grouped into 2 zones because of their distinct characteristics: rooms in zone Northwest are offices with windows, whereas zone Northeast includes elevators, restrooms and utility rooms, and does not have windows. 

This building is equipped with a variable air volume (VAV) HVAC system, that is common to 30\% of all U.S. commercial buildings \cite{VAV}. The system contains large supply fans that drive air over cooling coils, cooling it down to a desired supply air temperature, and then distribute the air to VAV boxes that govern the airflow to different building zones. The supply air may be reheated at the VAV box before entering the room. The 4th floor of SDH is served by 21 VAV boxes.

\subsection{Excitation Experiments and Data}

Data was collected during 11 non-consecutive weeks between September 2014 and June 2015. This time span includes periods when the building was under normal operation as well as periods with excitation experiments. 
Recorded data points include room temperatures measured at all VAV boxes on the 4th floor of SDH, air inflow rates from each VAV box, HVAC system's supply air temperature, outside ambient air temperature and solar radiation data recorded from a nearby weather station \cite{SolarRad}.

For accurate parameter identification, temperatures of neighboring zones should not be strongly correlated \cite{Lin_multizone}. 
For buildings in regular operation, this is generally achievable through forced response experiments. 
Because of commercial buildings' large thermal inertia, each forced excitation must last sufficiently long before temperature changes are observed. 
With these points in mind, we conducted our experiment as follows: Starting at 8am, every 2 hours, the supply airflow rate to one zone is set to its maximum value, minimum airflow rates are set for each of its neighboring zones and  a random airflow rate is chosen for each remaining zone. This is repeated for each of the 6 zones. This experiment is performed during weekends as (a) it minimizes effects due to building occupancy on our data, and thus the subsequent parameter identification; (b) temporary violation of comfort constraints during the weekend was allowed.

\section{Building Model}\label{sec:model}

\subsection{RC Modeling and the BRCM Toolbox}
We derive a Resistance-Capacitance (RC) model for our building using the Building Resistance-Capacitance Modeling (BRCM) MATLAB toolbox \cite{David}. 
The RC modeling methodology first decomposes a building into building elements (BE), such as the bulk volume of air in each room, walls, floors and ceilings. 
Then, an electric analogy is used to derive an equivalent electrical circuit whose resistances and capacitances represent thermal resistances and thermal capacitances of the BEs, and voltages and currents represent temperatures of BEs and heat transfers between those. The thermal resistances and capacitances of BEs are completely characterized by their geometry and construction data such as density, convection coefficient and specific heat capacity. 

In the BRCM toolbox, a building model consists of two parts: a thermal submodel and external heat flux submodels (EHFM). The thermal submodel describes passive heat transfer between the BEs and the EHFMs capture heat gain or loss due to external inputs and disturbances such as the outside environment. 
The BRCM toolbox semi-automates the derivation of an RC model by automatically computing the thermal submodel using the electrical circuit analogy and an input file that contains geometry and construction data of all BEs (e.g. we use an EnergyPlus file developed for the 4th floor of SDH as our input file). 
The EHFMs can be user defined, as different buildings may be subject to distinct inputs and disturbances.

\subsection{Building Model}

In this section, we first describe the EHFMs for our building and then present the final state-space model of the building.

There are 3 EHFMs:
\begin{itemize}
\item Building hull: convective heat transfer and solar radiation gains across exterior walls and windows.
\item HVAC: heat gain from the HVAC system.
\item Internal gains (IG): heat gain due to occupancy, electrical appliances and other unmodeled disturbances.
\end{itemize}

\noindent
Let $x \in \mathbb{R}^{289}$ be the state vector whose elements are the temperatures of all BEs on the 4th floor of SDH\footnote{In the EnergyPlus file for the 4th floor of SDH, each wall, floor and ceiling is decomposed into 2 to 3 BEs. In RC building models, the temperature of each BE is modeled by one state variable, thus the model of the 4th floor of SDH has a large number of states: 289.}, $u \in \mathbb{R}^{21}$ be the air inflow rate from the 21 VAV boxes on this floor and $v := \begin{bmatrix} v_\text{Ta}; v_\text{Ts}; v_\text{solE}; v_\text{solS}; v_\text{solW}; v_\text{solN} \end{bmatrix} \in \mathbb{R}^6$ be the disturbance vector whose elements represent ambient air temperature, supply air temperature from the HVAC system and solar radiation from the four geographical directions, respectively.

\textbf{Building Hull:~}
If the $i$-th BE is connected to the building hull, e.g. a room adjacent to the exterior wall, then the external heat fluxes acting on it due to the outside environment are modeled as:
\begin{equation}\label{eq:q_bh}
\begin{aligned}
q_{\text{BH},i} &= \gamma_{\text{EW}} \, a_{\text{EW},i} \big(v_\text{T\textsubscript{a}}(t) - x_{i}(t)\big) + \gamma_\text{absorp} \, a_{\text{EW},i} \,v_{\text{sol},i}(t) \\
	& \quad + U_{\text{win}} \, a_{\text{win},i} \big( v_\text{T\textsubscript{a}}(t) - x_{i}(t) \big) \\
	& \quad + \gamma_\text{winSolAbs} \, a_{\text{win},i} \, v_{\text{sol},i}(t),
\end{aligned}
\end{equation}
where $a_{\text{EW},i}$ and $a_{\text{win},i}$ are the total areas of the exterior wall and the window respectively, and $v_{\text{sol},i} \in \{v_\text{solE}; v_\text{solS}; v_\text{solW}; v_\text{solN}\}$ is the solar radiation affecting this BE. 
$\gamma_{\text{EW}}$, $\gamma_\text{absorp}$, $U_{\text{win}}$ and $\gamma_\text{winSolAbs}$ are tuning parameters of the model and their descriptions are given in Table \ref{tab:param}.

\textbf{HVAC:~}
If the $i$-th BE is a room equipped with at least one VAV box, then the heat flux acting on it is:
\begin{equation} \label{eq:q_hvac}
q_{\text{HVAC},i} = c_p \big(\textstyle \sum_{j \in \mathcal{B}_i} u_j(t) \big) \cdot \big(v_\text{T\textsubscript{s}}(t) - x_{i}(t)\big) ,
\end{equation}
where $c_p$ is the specific heat capacity of air, $\mathcal{B}_i$ is the set of VAV boxes serving the $i$-th room and $u_j$ is the $j$-th element of $u$. Note that due to the lack of temperature measurements of the supply air at the outlet of VAV boxes, $v_\text{Ts}$ is the supply air temperature upstream of the VAV boxes' heating coils, i.e., heat gains due to reheating at the VAV boxes are not modeled by (\ref{eq:q_hvac}), but are captured by the internal gains EHFM in our model. 

\textbf{Internal Gains:~}
If the $i$-th BE is a room, then it is also subject to internal gains, which are modeled as:
\begin{equation}\label{eq:q_ig}
q_{\text{IG},i} = a_{\text{floor},i} \big(c_{\text{IG},i} + f_{\text{IG},i}(t)\big),
\end{equation}
where $a_{\text{floor},i}$ is the room's floor area. 
$c_\text{IG} \in \mathbb{R}^6$ is an unknown constant vector representing a background time-invariant heat gain per unit area in each of the 6 zones, due to idle appliances. 
The function $f_\text{IG}(t): \mathbb{R} \rightarrow \mathbb{R}^6$ is an unknown function that captures time-varying internal gains in different zones. 
Finally, $c_{\text{IG},i}$ and $f_{\text{IG},i}(t)$ are the relevant elements of $c_\text{IG}$ and $f_\text{IG}(t)$ that correspond to the $i$-th room.

After defining all EHFMs, the BRCM toolbox automatically generates the following model:
\begin{equation} \label{eq:ct_model}
\begin{aligned}
\dot x(t) &= A_t x(t)+B_t \big[ q_\text{BH} \big(x(t),v(t) \big) \\
	& \quad + q_\text{HVAC}\big( x(t), v(t), u(t) \big) + q_\text{IG}(t) \big], \\
	& = Ax(t) + B_v v(t) + B_\text{IG} \big(c_\text{IG} + f_\text{IG}(t) \big) \\
	& \quad + \textstyle \sum_{j=1}^{21} \big( B_{xu_j} x(t) + B_{vu_j} v(t) \big) u_j(t)
\end{aligned}
\end{equation}
where in the first equality, $A_t$, $B_t$ represent the thermal submodel, and $q_\text{BH}(\cdot), ~q_\text{HVAC}(\cdot), ~q_\text{IG}(\cdot)$ are the EHFMs, which are vector-valued functions as follows:
if the $i$-th BE is not subject to a specific EHFM, say HVAC, then $q_{\text{HVAC},i} = 0$, otherwise $q_{\text{HVAC},i}$ is given by Equation (\ref{eq:q_hvac}). The second equality is obtained by expressing  $q_\text{BH}, ~q_\text{HVAC}$ and $q_\text{IG}$ as functions of $x$, $u$ and $v$, using (\ref{eq:q_bh}) to (\ref{eq:q_ig}). The bilinearities in (\ref{eq:ct_model}) naturally arise from the physics of the HVAC system (refer to Equation (\ref{eq:q_hvac})).

Finally, we discretized the model using a fixed time step of 15 min to obtain its approximate discrete time model, which is semi-parametric and bilinear:
\begin{equation}\label{eq:model_dt}
\begin{aligned}
x(k+1) &= Ax(k) + B_v v(k) + B_\text{IG} \big(c_\text{IG} + f_\text{IG}(k) \big) \\
	& \quad + \textstyle \sum_{j=1}^{21} \big( B_{xu_j} x(k) + B_{vu_j} v(k) \big) u_j(k) \\
y(k) &= C x(k),
\end{aligned}
\end{equation}
where $x$, $u$ and $v$ are as defined before, and $y \in \mathbb{R}^6$ represents the measured average temperature of each zone.

\begin{table}
	\caption{Model parameters.}
	\centering
	\begin{tabular}{l l l}
	\hline
	Parameter & Description & Value [Unit]\\
	\hline
	$\gamma_{\text{EW}}$ & exterior wall convection coeff. & 10.5  [W/(m$^2$K)]\\
	$\gamma_{\text{IW}}$ & interior wall convection coeff. & 29.4  [W/(m$^2$K)]\\
	$\gamma_{\text{floor}}$ & floor convection coeff. & 51.5  [W/(m$^2$K)]\\
	$\gamma_{\text{ceil}}$ & ceiling convection coeff. & 44.3  [W/(m$^2$K)]\\
	$\gamma_{\text{absorp}}$ & ext. wall solar absorption coeff. & 0.75  [-]\\
	$\gamma_{\text{winSolAbs}}$ & window solar absorption coeff. & 0.03  [-]\\
	$U_{\text{win}}$ & window heat transmission coeff. & 0.63  [W/(m$^2$K)]\\
	$c_{\text{IG,NW}}$ & background heat gain in zone NW & 0.3  [W/m$^2$]\\
	$c_{\text{IG,W}}$ & background heat gain in zone W & 8.0  [W/m$^2$]\\
	$c_{\text{IG,S}}$ & background heat gain in zone S & 18.8  [W/m$^2$]\\
	$c_{\text{IG,E}}$ & background heat gain in zone E & 8.0  [W/m$^2$]\\
	$c_{\text{IG,NE}}$ & background heat gain in zone NE & 11.0  [W/m$^2$]\\
	$c_{\text{IG,C}}$ & background heat gain in zone C & 8.0  [W/m$^2$]\\	
	\hline
	\end{tabular}	
	\label{tab:param}
\end{table}

%% file: results.tex
\section{Model Identification}\label{sec:model_id}

The model identification process is carried out in two steps.
In Section \ref{sec:param_id}, we identify the model parameters listed in Table \ref{tab:param}. 
To simplify the parameter identification process, we use the approximation $f_\text{IG}(k) = 0$ during weekend days, in order to reduce (\ref{eq:model_dt}) to a purely parametric model. 
With the optimal parameter values in hand, we then identify the function $f_\text{IG}(\cdot)$ in Sections \ref{sec:fixed_ig}.

\subsection{Parameter Identification}\label{sec:param_id}

The model parameters are estimated using data collected during two weekends in spring and summer, when excitation experiments were carried out. The identified model is then validated on data collected during a weekend in fall (using $f_\text{IG}(k) = 0$).

Let $\gamma \in \mathbb{R}^{13}$ be the parameter vector whose elements are those parameters listed in Table \ref{tab:param}. We choose the optimal $\hat \gamma$ that solves the following optimization problem:
\begin{equation}
\begin{aligned}
 \hat{\gamma} =&~\arg\min~\textstyle \sum_k \Vert y(k,\gamma) - \bar y(k) \Vert ^ 2 \\
\text{s.t.~~}
&y(k,\gamma) \text{~and~} x(k,\gamma) \text{~satisfy (\ref{eq:model_dt}) with } f_\text{IG}(k) = 0 ~\forall k\\
&u(k) = \bar u(k), ~ v(k) = \bar v(k)~\forall k\\
&x(0) = x_{\text{KF}}(0)\\
&\gamma > 0, ~0 \leq \gamma_\text{absorp}, \gamma_\text{winSolAbs} \leq 1,
\end{aligned}
\end{equation}
\noindent
where $\bar y$, $\bar u$ and $\bar v$ are measured zone temperatures, inputs and disturbances, respectively; and $x_{\text{KF}}(0)$ is the initial state estimated using a Kalman Filter. Estimating $x(0)$ is necessary since not all states can be measured (for example, the wall temperatures are not). 
In other words, we choose the set of parameters such that when the model is simulated with this set of parameters and the measured inputs and disturbances, the sum of squared errors between the measured temperatures and the simulated temperatures is minimized. 
We can use prior knowledge about the building to compensate for limited excitation of the system, e.g. we use initial guesses for parameter values that are physically plausible.

The identified parameter values are reported in Table \ref{tab:param}. 
The root-mean-square (RMS) errors between the model's simulated temperature for different zones and the measured temperatures for the training data are shown in Table \ref{tab:rms_param}.
This table also shows the RMS prediction errors when the identified model is used to predict a validation dataset (also on weekend data, using $f_\text{IG}(k) = 0$).

\begin{table}
\caption{RMS error from parameter identification using weekend data,  by zone, for the training and validation datasets.}
\centering
\begin{tabular}{*8c}
\toprule
& NW & W & S & E & NE & C & Mean \\ 
\midrule
Training & 0.46 & 0.41 & 0.24 & 0.34 & 0.27 & 0.28 & 0.333\\
\midrule
Validation & 0.62 & 0.57 & 0.31 & 0.28 & 0.39 & 0.31 & 0.413\\
\bottomrule
\end{tabular}
\label{tab:rms_param}
\end{table}

\subsection{Identification of the Time-Varying Internal Gains}\label{sec:fixed_ig}

A random subset of 8 weeks is selected from the entire dataset and used as training data for estimating the time-varying internal gains function $f_\text{IG}(\cdot)$, and the remaining 3 weeks of data are used as a validation set. 
For each week $w$ in the training set, we estimate an instance of this function, $f_{\text{IG},w}(\cdot)$. The final estimate of the function $f_\text{IG}(\cdot)$ is defined as the average of all estimates $f_{\text{IG},w}(\cdot)$.

More specifically, 
at each time $k$, let $\tilde x_w(k)$ and $\tilde y_w(k)$ denote the simulated state and measurement vectors with $f_{\text{IG},w}(k-1) = 0$, i.e.,:
\begin{equation}\label{Eq:xnoig_ynoig}
	\begin{aligned}
	\tilde x_w(k) & = Ax_w(k-1)+B_v v_w(k-1) + B_\text{IG} c_\text{IG} \\
		& \quad + \textstyle \sum_{j=1}^{21} \big( B_{xu_j} x_w(k-1) + B_{vu_j} v_w(k-1) \big) \\
		& \quad \cdot u_{w,j}(k-1) \\
	\tilde y_w & = C \tilde x_w(k).
	\end{aligned}
\end{equation}
By noting $x_w(k) = \tilde x_w(k) + B_\text{IG} f_{\text{IG},w}(k-1)$, we can estimate $f_{\text{IG},w}(k-1)$ by solving the following set of linear equations using ordinary least-squares:
\begin{equation}\label{Eq:ig(k-1)}
(C B_\text{IG}) \cdot f_{\text{IG},w}(k-1) = \bar y_w(k) - \tilde y_w(k),
\end{equation}
where $\bar y_w(k)$ is the measured zone temperature at time $k$ from the $w$-th training week. 
Finally, the estimate $\hat f_\text{IG}(\cdot)$ is obtained by: 
\begin{equation}\label{Eq:fixed_ig}
\hat f_\text{IG}(k) = \frac{1}{8} \textstyle \sum_{w=1}^{8} f_{\text{IG},w}(k) \quad \text{for all }k.
\end{equation}

\subsection{Impact of Internal Gains}

\begin{figure}
	\center
	\includegraphics[width=1\columnwidth]{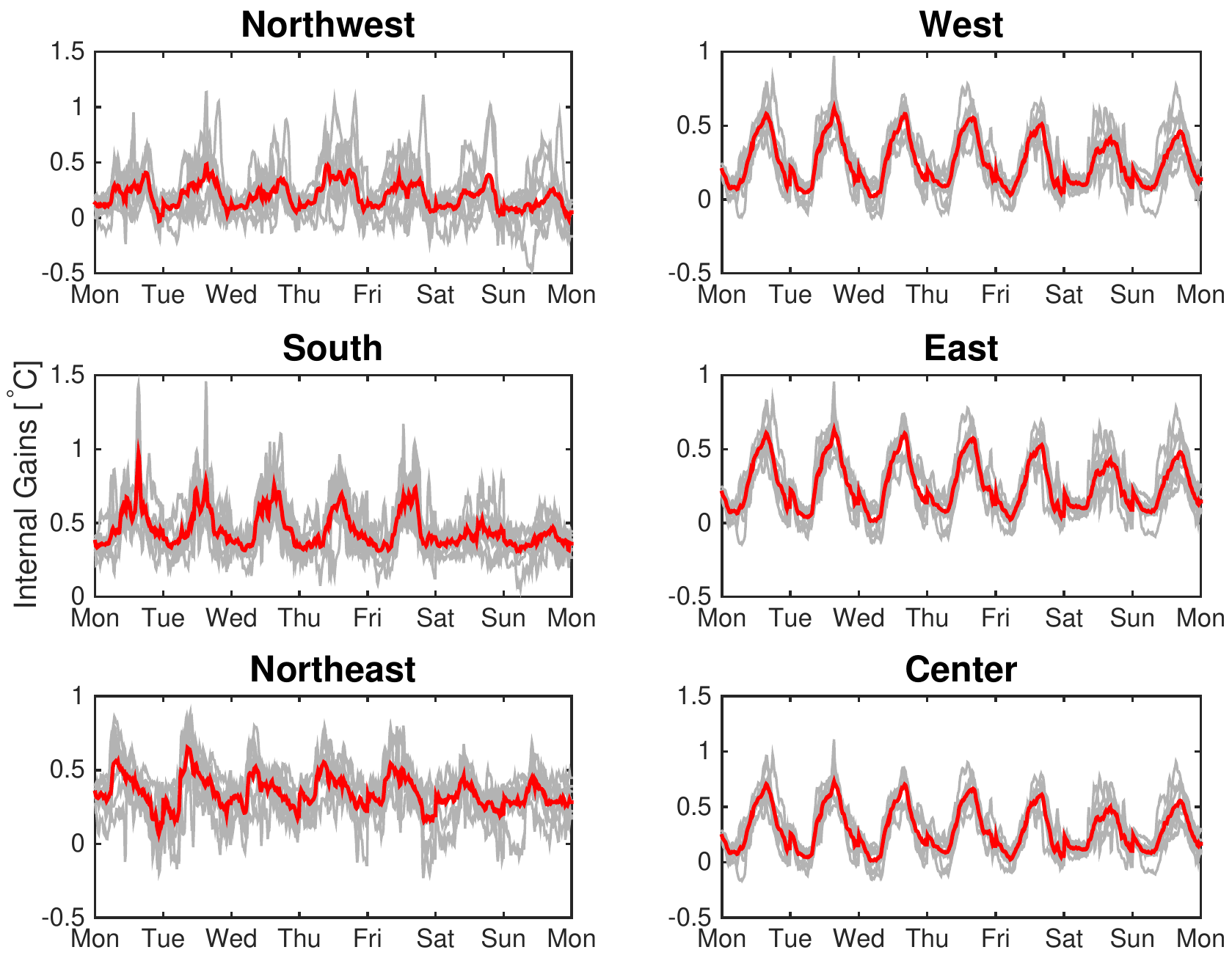}
	\caption{Estimated increase in temperature due to internal gains by zone. Gray lines are estimates for individual weeks in the training dataset, red lines are their averages.}
	\label{fig:ig}
\end{figure}

The estimated average increase in room temperature due to internal gains, i.e., $B_\text{IG} \big( c_\text{IG} + \hat f_\text{IG}(k) \big)$ is shown in red in Figure \ref{fig:ig}.
It can be observed that internal gains profiles vary on both long and short time horizons from approximately 0\degree C to 1\degree C. A slightly larger increase in temperature of approximately 1.4\degree C is reported in \cite{Aswani} for a similar office space. This may be because the internal gains term in their model also includes heat gain from solar radiation, whereas our model captures the effects of solar radiation separately. 
Observe that the internal gains profiles increase during the day, peaking in the early afternoon and then slowly decrease, reaching a minimum at night. 
In addition, the profiles' peaks are slightly lower during the weekends. 
These patterns coincide with when building occupants typically come into and leave the office. 

The gray lines in Figure \ref{fig:ig} show the same quantity estimated for each training week, i.e., $B_\text{IG} \big( c_\text{IG} + f_{\text{IG},w}(k) \big)$. 
It can be observed that different zones experience different variations in internal gains across the training weeks. 
The zones West, East and Center are workspaces of students who tend to have regular schedules and hence, more regular internal gains patterns. 
On the other hand, the remaining three zones experience more uncertainty in internal gains, possibly due to the presence of windows, elevators and staircases, and known inaccuracies in the EnergyPlus input file for zone Northwest.

\begin{figure}
	\center
	\includegraphics[width=1\columnwidth]{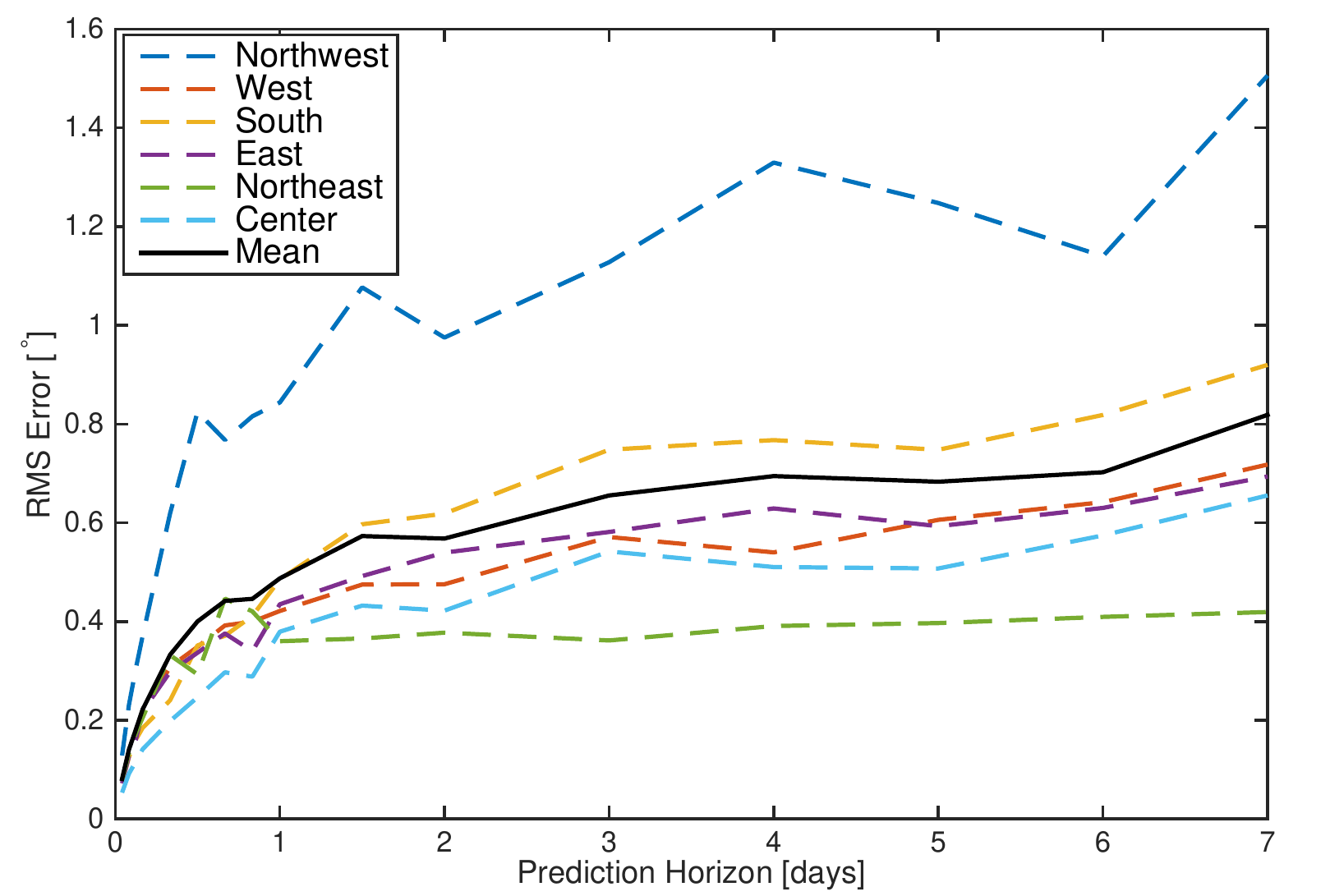}
	\vspace*{-0.6cm}
	\caption{RMS prediction error versus prediction horizon.}
	\label{fig:pred_horizon}
\end{figure}

\begin{figure}
	\center
	\includegraphics[width=1\columnwidth]{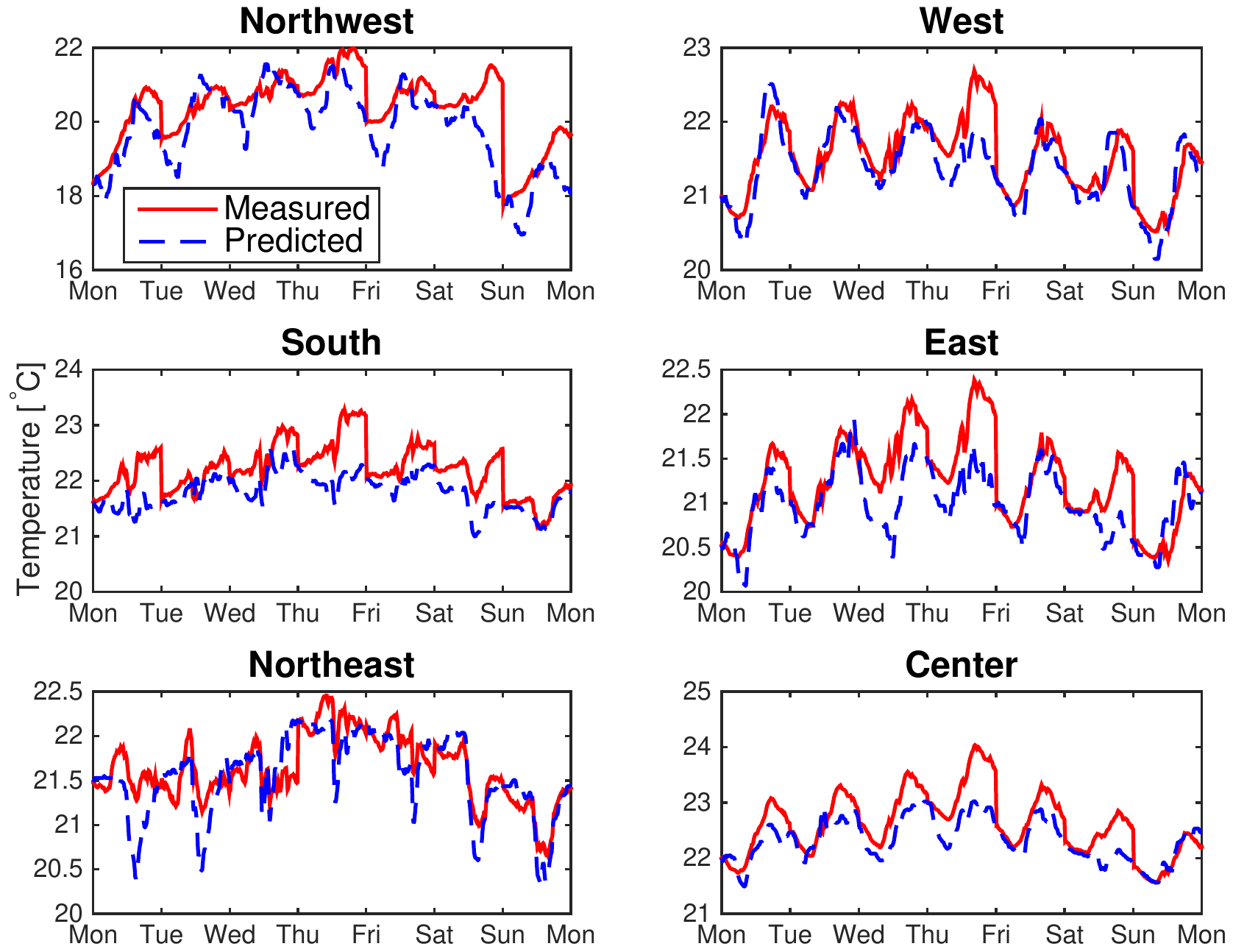}
	\vspace*{-0.6cm}
	\caption{Predicted temperature using fixed internal gains function, by zone.}
	\label{fig:fixed_ig}
\end{figure}

The identified model with $\hat f_\text{IG}$ (a fixed function) is used to make 24-hour predictions of zone temperatures for all 3 weeks in the validation set, i.e., the state vector is estimated by a Kalman Filter every 24 hours.
Figure \ref{fig:fixed_ig} shows the results for one of these weeks and the average RMS prediction error for all validation weeks is reported in Table \ref{tab:rms_ig}. Furthermore, Figure \ref{fig:pred_horizon} shows that the model's prediction accuracy decreases with increasing prediction horizon, which could be explained by uncertainties in internal gains as well as model inaccuracies. It is interesting to observe that the RMS prediction error for zone Northwest is the largest and it also increases the fastest as the prediction horizon increases, which is in accordance with this zone experiencing large variations in internal gains (Figure \ref{fig:ig}) and its geometry data in the EnergyPlus input file being inaccurate.

\section{Online Update of the Internal Gains Function} \label{sec:online_ig}

The previous section shows that the large time-varying internal gains are difficult to capture \textit{a priori}, nevertheless, they can significantly affect our model's prediction accuracy. 
In light of this, we consider a learning based approach in this section, where we update the internal gains function $f_\text{IG}(\cdot)$ online using past observations.

In other words, instead of estimating a fixed function $\hat f_\text{IG}(\cdot)$ \textit{a priori}, at every time $k$, we estimate $f_\text{IG}(k-1)$ using (\ref{Eq:xnoig_ynoig}) and (\ref{Eq:ig(k-1)}), and then use the following simple model
\begin{equation}\label{eq:online_ig}
f_\text{IG}(k) = f_\text{IG}(k-1)
\end{equation}
to obtain an online estimate of $f_\text{IG}(k)$, which is then used in (\ref{eq:model_dt}) to predict $x(k+1)$ and $y(k+1)$. 
The intuition for (\ref{eq:online_ig}) is that internal gains do not change significantly from time $k-1$ to $k$ (i.e., 15 min). 

\begin{figure} 
	\center
	\includegraphics[width=1\columnwidth]{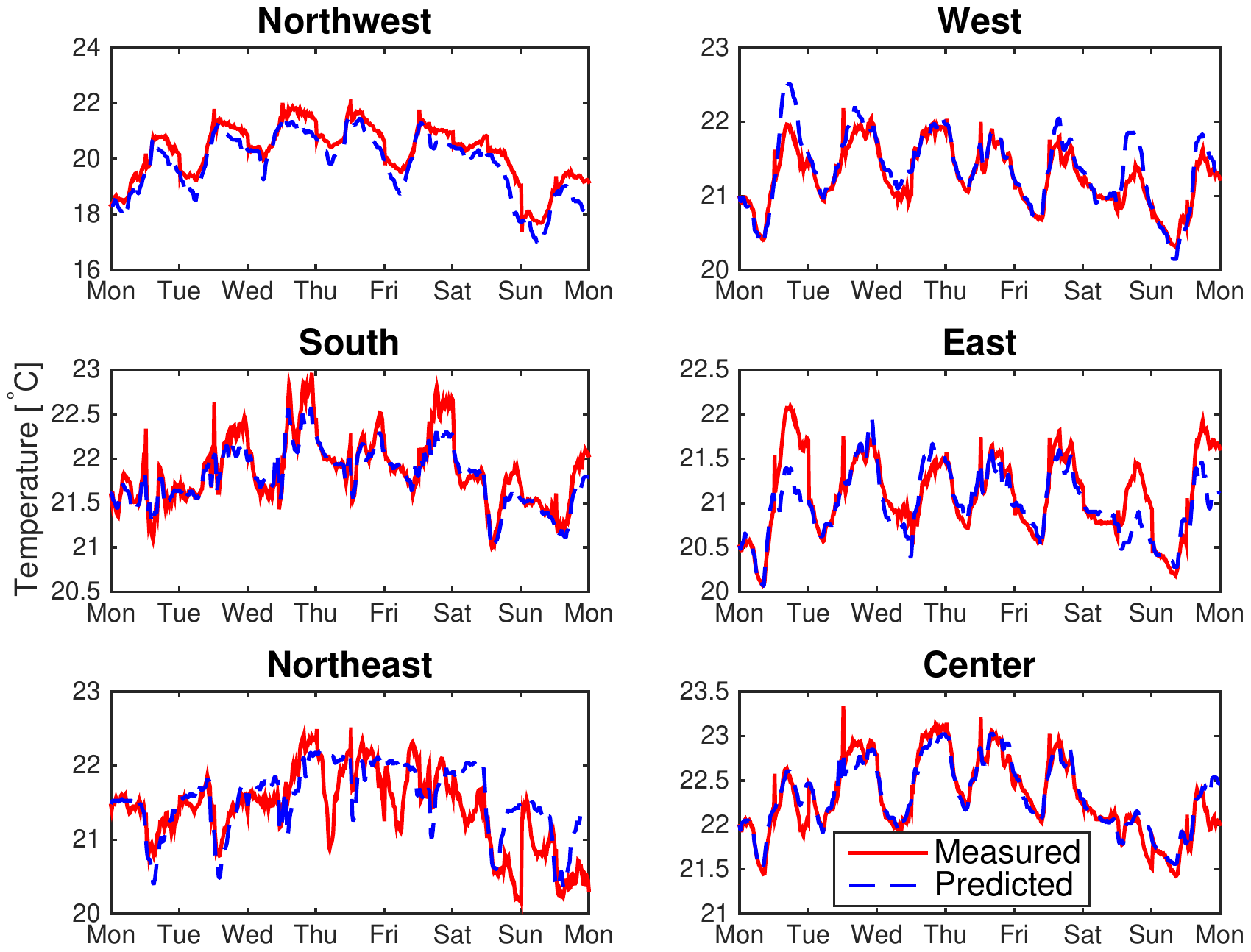}
	\vspace*{-0.6cm}
	\caption{Predicted temperature using online updated internal gains function, by zone.}
	\label{fig:online_ig}
\end{figure}

This model is simulated for all 11 weeks of data, one of which is shown in Figure \ref{fig:online_ig}. The average RMS prediction errors are reported in Table \ref{tab:rms_ig}. Thus, by dynamically updating $f_\text{IG}(\cdot)$ online, the model's prediction accuracy is improved by 36\% on average compared with when a fixed internal gains function was used.

\begin{table}[hbtp]
\caption{RMS prediction error by zone for a model with fixed $f_\text{IG}$ and using online updated $f_\text{IG}$.}
\centering
\begin{tabular}{*8c}
\toprule
& NW & W & S & E & NE & C & Mean \\ 
\midrule
Fixed $f_\text{IG}$ & 0.84 & 0.42 & 0.48 & 0.43 & 0.36 & 0.38 & 0.485\\
\midrule
Online Updated \\$f_\text{IG}$ & 0.50 & 0.31 & 0.15 & 0.41 & 0.32 & 0.16 & 0.308\\
\bottomrule
\end{tabular}
\label{tab:rms_ig}
\end{table}


\section{Discussion}\label{Sec:discussion}

In Section \ref{sec:model_id}, we conducted excitation experiments to actively perturb our building. 
Data collected during the experiments and additional weekends is used with the approximation that the time-varying internal gains are zero, to identify the model parameters. Then, we estimate a fixed internal gains function, $f_\text{IG}(\cdot)$, using 8 weeks of measurements. The resulting model is used to make 24-hour predictions of the building's temperature profile for 3 additional weeks, and an average RMS error of 0.48\degree C is achieved. 
Figures \ref{fig:ig} and \ref{fig:pred_horizon} suggest that our building is subject to large uncertain internal gains which make accurate long term predictions difficult. 
For buildings that are subject to fewer uncertainties, a model identified using this procedure may achieve better prediction accuracy. 

In addition, there are several approaches that can be taken to further enhance the model's prediction accuracy. 
When weekend data is used to identify the model parameters, more sophisticated approximations of the occupancy function, such as sinusoids, can be used. 
Sensors can be installed in the VAV boxes to measure the temperature of the supply air, from the HVAC system, downstream of the heating coils. 
Moreover, occupancy sensors can be used to improve the estimate of internal gains and hence the model's prediction quality.

In Section \ref{sec:online_ig}, we dynamically updated our estimate of the internal gains function using current temperature measurements. 
More specifically, we assume the current heating load, due to internal gains, remains constant during the next time step. We apply this model to make 24-hr predictions and demonstrate that its prediction accuracy is significantly improved (compare Figure \ref{fig:fixed_ig} with Figure \ref{fig:online_ig}). 
In addition, using more sophisticated regression techniques 
and taking into account other factors such as past heating loads and room temperatures may further improve the prediction accuracy and extend the prediction horizon. 

For the frequency regulation application, a model is first applied to estimate the building's power consumption for the next 24 hours, in order to determine its reserve capacity for the day-ahead reserve market. MPC can then be used to provide these reserves without violating comfort constraints. Thus, by learning $f_\text{IG}(\cdot)$ online and dynamically updating it in an MPC controller, errors from the 24-hour prediction may be corrected during reserve provision.